\documentclass[final,5p,times, twocolumn]{elsarticle}
\usepackage{graphicx}
\usepackage{siunitx}
\usepackage{color}
\usepackage{amssymb}
\usepackage{amsmath}
\usepackage{xspace}
\usepackage{float}
\usepackage{multirow}   
\usepackage{booktabs}   
\usepackage{longtable}
\usepackage{algorithm}
\usepackage{algorithmic}
\usepackage{upgreek}
\usepackage{xfrac}

\usepackage{hyperref}
\hypersetup{colorlinks=true, linkcolor=black, citecolor=black, urlcolor=black}
\usepackage{lineno}
\graphicspath{{figures/}}

\journal{Soft Matter}
\begin{document}
\begin{frontmatter}



\title{Modeling liquid-mediated interactions for close-to-substrate magnetic microparticle transport in dynamic magnetic field landscapes}
\author[a]{Markus Gusenbauer\corref{cor1}}
\ead{markus.gusenbauer@donau-uni.ac.at}
\cortext[cor1]{Corresponding authors}
\author[b]{Rico Huhnstock\corref{cor1}}
\ead{rico.huhnstock@physik.uni-kassel.de}
\author[a]{Alexander Kovacs}
\author[a]{Harald Oezelt}
\author[b]{Arno Ehresmann}
\author[a,c]{Thomas Schrefl}
\address[a]{Department for Integrated Sensor Systems, University for Continuing Education Krems, 3500 Krems, Austria}
\address[b]{Institute of Physics and Center for Interdisciplinary Nanostructure Science and Technology (CINSaT), University
of Kassel, 34132 Kassel, Germany.}
\address[c]{Christian Doppler Laboratory for Magnet design through physics informed machine learning, 2700 Wiener Neustadt, Austria}


\begin{abstract}
Understanding the on-chip motion of magnetic particles in a microfluidic environment is key to realizing magnetic particle-based Lab-on-a-chip systems for medical diagnostics. In this work, a simulation model is established to quantify the trajectory of a single particle moving close to a polymer surface in a quiescent liquid. The simulations include hydrodynamic, magnetostatic, and Derjaguin-Landau-Verwey-Overbeek (DLVO) interactions. They are applied to particle motion driven by a dynamically changing magnetic field landscape created by engineered parallel-stripe magnetic domains superposed by a homogeneous, time-varying external magnetic field. The simulation model is adapted to experiments in terms of fluid-particle interactions with the magnetic field landscape approximated by analytic equations under the assumption of surface charges. Varying simulation parameters, we especially clarify the impact of liquid-mediated DLVO interactions, which are essential for diagnostic applications, on the 3D trajectory of the particle. A comparison to experimental results validates our simulation approach.
\end{abstract}

\end{frontmatter}


\section{\label{sec:introduction}Introduction}
The actuation of magnetic particles in a microfluidic environment via magnetic fields has high application potential in biosensing Lab-on-a-chip systems~\cite{Gijs2004,Vanreenen2014}. Due to their large surface-to-volume ratio and flexibility in chemical surface properties, magnetic particles are already playing a significant role in lab-based bio-detection assays~\cite{Vanreenen2014}. They are commercially available and come in different sizes, surface functionalizations, and magnetic properties~\cite{Vanreenen2014}. When implemented in Lab-on-a-chip devices, magnetic particles move close to an underlying substrate surface, immersed in a fluid. An established scheme for achieving fast and robust on-chip particle actuation is magnetophoresis by dynamically superposing magnetic stray field landscapes above micromagnetic structures with varying external homogenous magnetic fields~\cite{Rampini2016,abedininassab2023} (sometimes also termed traveling wave magnetophoresis~\cite{Yellen2007}). These structures may either be arrays of specifically shaped magnets~\cite{Block2023,Rampini2021,Sadeghidelouei2024}, magnetic domains forming naturally~\cite{Tierno2007}, or imprinted artificially in topographically flat magnetic thin films~\cite{Ehresmann2015}.

Predicting the particles' trajectories has in the past been carried out by estimating the equilibrium distance between particle and substrate surface due to the force balance between the liquid-mediated surface-surface Derjaguin, Landau, Verwej, Overbeek (DLVO) forces~\cite{Verwey1947} and the (out-of-plane) $z$-component of the magnetic force, solving the equation of motion for a planar trajectory at this distance~\cite{Wirix-Speetjens2005}.

In a recent attempt also the motion of a particle in the third dimension has been included in trajectory simulations, predicting in addition to the lateral motion also vertical upward jumps of the particles~\cite{Klingbeil2020}, which have been experimentally quantified for a specific magnetic stray field landscape
using a defocusing-based 3D particle tracking method~\cite{huhnstock2022three}. As it turns out, particles are jumping from one force equilibrium position (e.g., close to a domain wall) to another directly after inverting the direction of the external field. The upward movement is due to an inverted magnetic force repelling the particle from the substrate surface. Once the particle is approaching the adjacent force equilibrium position as a result of its lateral motion, the magnetic force turns attractive again, pulling the particle towards the substrate surface and concluding the jump. Little attention has been paid to the contribution of the liquid-mediated particle-substrate interaction as the influence of the underlying substrate has been accounted for by only including an empirical $z$-dependent drag coefficient, leading to increased drag when the particle is closer to the substrate surface~\cite{Wirix-Speetjens2005,Klingbeil2020}. However, if two solids surrounded by a fluid are close to each other, DLVO surface forces must be taken into account. 
DLVO interactions depend on the surface potentials of the particle and the substrate and on the ionic strength of the surrounding liquid. They are, therefore, important 
for biosensing devices where induced binding of an analyte, e.g., a protein, changes surface potentials and, by extension, the DLVO interaction. This has direct consequences for the 3D magnetic particle motion in a dynamic magnetic stray field landscape; 
measuring respective changes for the particles' 3D-trajectories may, therefore, indicate analyte binding events.

\begin{figure}
    \centering
    \includegraphics[width=\linewidth]{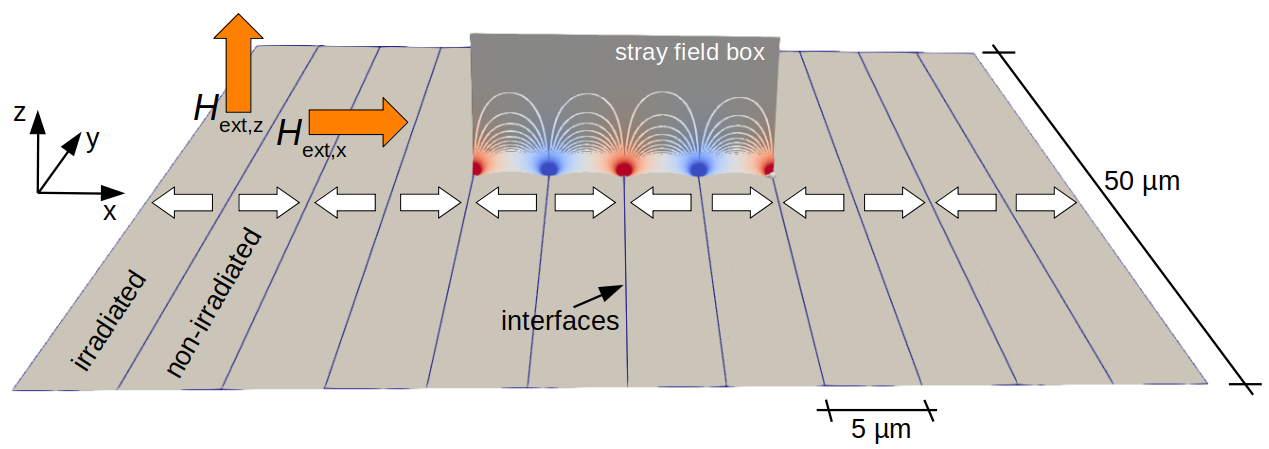}
    \caption{Model of the bilayer system with an analytic stray field box. Only the ferromagnetic Co$_{70}$Fe$_{30}$ layer is visible. The external field is applied in $z$- and $x$-direction according to Fig.~\ref{fig:start_simulation}. The setup consists of multiple stripes in $x$-direction with periodically repeated head-to-head/tail-to-tail magnetization configuration, created by ion bombardment induced magnetic patterning. Typical stray field lines are given in arbitrary values from positive (red) to negative (blue) $z$-field component.}
    \label{fig:model}
\end{figure}

We seek in this work a thorough understanding of the multiparameter space that impacts the close-to-substrate 3D trajectory of a single micron-sized particle in a dynamically transformed stray field landscape, stemming from a head-to-head/tail-to-tail magnetized parallel-stripe domain pattern of a topographically flat magnetic layer system (Fig.~\ref{fig:model}). 

In the specific example considered here, the particle is moving from a head-to-head domain wall to a tail-to-tail domain wall location upon quickly reversing the external magnetic field component perpendicular to the substrate surface ($z$-direction) while applying a constant field in the parallel-to-surface (+x)-direction (Fig.~\ref{fig:model}). Special attention is paid to the influence of DLVO forces on the particle's motion.

Previously, the motion of spherical particles in viscous fluids has been approximated by a Stokes drag for laminar flow~\cite{deymier1994molecular}. Alternatively, complex hydrodynamics and magnetic particle interactions have been solved by finite element simulations using Navier-Stokes and continuous Maxwell equations~\cite{ly1999simulations}, yet still, the interaction of particles with the fluid has also been approximated by the Stokes drag. The development of numerical methods to describe magnetic particles in fluids is strongly influenced by the research of magnetorheological fluids, either by continuous or discrete approaches~\cite{ghaffari2015review}. 

Here, we are interested in describing magnetic particle motion accounting for full hydrodynamic interactions close to a flat surface. In a previous work, we computed magnetic particles in discretized lattice Boltzmann fluid dynamics~\cite{gusenbauer2018simulation}, which is implemented in the soft-matter simulation environment ESPResSo~\cite{weik2019espresso}. In this past work, the particles were magnetically saturated by a strong external field. In the present work, in contrast, the particle is superparamagnetic and not fully saturated. The working point lies, in our case, on the ascending slope of the particle's magnetization curve. Particle and fluid are treated explicitly. Mass, volume, magnetic susceptibility, and magnetization act directly in the center of the particle, which moves in the effective field resulting from the vector sum of the magnetic stray field landscape and the external field. DLVO forces act at the particle's surface, depending on adhesion and electrical charge properties of both particle and substrate surfaces. Gravitation and buoyancy forces act in the center of the particle as well. In the simulation, the particle interacts indirectly with the surrounding fluid, where fictive fluid molecules perform consecutive propagation and collision processes. The movement of the particle acts on the fictive fluid molecules and vice versa for a full two-way hydrodynamic interaction.

Our proposed simulation model includes a large number of parameters, some of which are very sensitive to small changes. However, not all parameters can be experimentally quantified to the precision required for tuning the computational model. In this work, we probe the sensitivity of a particle's 3D trajectory to changes in these parameters. The results will allow a better understanding and control of particle trajectories in experiments and Lab-on-a-chip devices.

\section{\label{sec:methods}Methods}
\subsection{Experimental setup\label{ssec:methods_experiment}}
For tracking the 3D motion of magnetic particles, an optical defocusing-based approach~\cite{huhnstock2022three} has been used. Briefly, the setup consists of an optical reflected-light microscope using a 100x Nikon-objective (leading to a small depth of focus), equipped with a movable sample stage surrounded by a coil arrangement providing the homogeneous magnetic field pulses in any direction and a high-speed video camera (Optronis CR450×2). The video camera has been operated with 1000 frames per second at 800 px $\times$ 600 px image resolution. Pulses in $x$- and $z$-directions (see Fig.~\ref{fig:model} for the used coordinate system) for inducing particle motion are trapezoidal with plateaus at \SI{1}{\milli\tesla} in $x$- and at $\pm$\SI{8}{\milli\tesla} in $z$-direction. For the motion step investigated in this work, the $z$-field switches from +\SI{8}{\milli\tesla} to -\SI{8}{\milli\tesla} within \SI{20}{\milli\second} at constant $x$-field.
\subsection{Microfluidic chip\label{ssec:methods_chip}}
The substrate of the microfluidic chip consists of an exchange bias thin film layer system starting with \SI{10}{\nano\meter} Cu as a buffer layer, followed by \SI{30}{\nano\meter} Ir$_{17}$Mn$_{83}$ as the antiferromagnetic layer, \SI{10}{\nano\meter} Co$_{70}$Fe$_{30}$ as the ferromagnetic layer and finished by \SI{10}{\nano\meter} Au as a capping layer, deposited via rf-sputtering on a naturally-oxidized Si wafer (\SI{1.5}{\centi\meter}~$\times~$\SI{1.5}{\centi\meter}). After deposition, the sample was treated with a field cooling procedure to initialize the exchange bias direction, heating it up to \SI{300}{\degreeCelsius} inside an in-plane magnetic field of \SI{145}{\milli\tesla} for \SI{60}{\minute}. The field-cooled sample was patterned into parallel-stripe domains via light-ion bombardment induced magnetic patterning~\cite{Mougin2001,Ehresmann2006}. It was coated with a resist layer structured into periodically repeated \SI{5}{\micro\meter} stripes using optical lithography. The thickness of the resist was chosen so that \SI{10}{\kilo\electronvolt} He ions do not penetrate through it. After treatment with \SI{10}{\kilo\electronvolt} He ion bombardment with a dose of $10^{15}$~ions/cm² inside an in-plane magnetic field (\SI{100}{\milli\tesla}) that is directed antiparallel to the initial exchange bias direction, the sample exhibits a periodic domain pattern with \SI{5}{\micro\meter} wide stripes and a head-to-head/tail-to-tail magnetization configuration (Fig.~\ref{fig:model}). The resist structure is then removed from the sample surface. A polymer layer, Poly(methyl methacrylate) (PMMA), with a thickness of \SI{150}{\nano\meter} was deposited on the sample surface via spin coating to serve as a spacer layer and prevent sticking of the particles to the metallic sample surface. A Parafilm with a window of \SI{8}{\milli\meter} $\times$ \SI{8}{\milli\meter} has then been adhered to the substrate, into which the solution with the magnetic particles has been pipetted. The container was then sealed with a glass cover slip.
\subsection{DLVO forces\label{ssec:methods_DLVO}}
DLVO forces account for the electrodynamic van-der-Waals interaction between two surfaces separated by a fluid and the electrostatic Coulomb interaction inherent to the surface adsorption of ions from the fluidic environment~\cite{Wirix-Speetjens2005}. The DLVO theory can be employed to characterize the van-der-Waals force $\vec F_{\text{vdW}}$ and the electrostatic force $\vec F_{\text{el}}$ acting between particle and substrate surfaces in the investigated system. The interaction energy between two bodies of arbitrary geometry is given by the interatomic van-der-Waals potential $w(r) = -C/r^6$~\cite{Israelachvili.2011}. 

In addition, the Hamaker constant $A_{\text{132}} = \pi^2  C  \rho_{\text{1}} \rho_{\text{2}}$ can be used to specify the interaction of the bodies~\cite{Israelachvili.2011}. $C$ describes an interaction parameter, while $\rho_{\text{1}}$ and $\rho_{\text{2}}$ are the number of atoms in each body, respectively. For a spherical particle of material 1 with radius $r$, which is positioned within a medium 3 in a distance $z$ from material 2 with an infinitely flat surface, the van-der-Waals force can be expressed as the following using an appropriate Hamaker constant $A_{\text{132}}$~\cite{Gregory1981,Wirix-Speetjens2005}:
\begin{equation}
\label{eq:vdw}
\vec{F}_{\text{vdW}}\left(z\right) = -\frac{A_{\text{132}}  r}{6  z^{2}}  \left(\frac{1}{1 + 14  z/\lambda_{\text{ret}}}\right)  \hat{e}_{\text{z}} .
\end{equation}
$\lambda_{\text{ret}}$ represents a characteristic retardation wavelength for the interaction and $\hat{e}_{\text{z}}$ stands for the unit vector in \textit{z}-direction.\\
The electrostatic interaction force $\vec{F}_{\text{el}}(z)$ between two surfaces immersed in a fluid is the consequence of the formation of an electrical double layer at each surface caused by the accumulation of positively and negatively charged ions from the fluid~\cite{Butt2003}. This double layer impacts the surface potentials $\Psi_{\text{p}}$ and $\Psi_{\text{s}}$ of a spherical particle and a flat surface, respectively. Attractive or repulsive electrostatic forces $\vec F_{\text{el}}(z)$ between particle and substrate surfaces can be calculated by considering the respective surface potentials~\cite{Wirix-Speetjens2005}:
\begin{align}
\begin{split}
\label{eq:el}
\vec{F}_{\text{el}}\left(z\right) = \frac{2  \pi  \epsilon  \kappa  r}{1-e^{-2  \kappa  z}}  \left[2  \Psi_{\text{s}}  \Psi_{\text{p}}\right. & e^{-\kappa  z} \\ \mp \left(\Psi_{\text{s}}^{2} + \Psi_{\text{p}}^{2}\right) &\left. e^{-2  \kappa  z}\right]  \hat{e}_{\text{z}} ,
\end{split}
\end{align}

with $\epsilon$ being the permittivity of the surrounding fluid medium. The Debye-Hückel inverse double layer thickness $\kappa$ is a function of the temperature $T$ and the ionic strength of the fluid $I$~\cite{Wirix-Speetjens2005}. It can be expressed by~\cite{Wirix-Speetjens2005}
\begin{equation}
\label{eq:kappa}
\kappa = \sqrt{\frac{2000  N_{\text{A}}  e^2 \cdot I}{\epsilon  k_{\text{B}}  T}} ,
\end{equation}
where $N_{\text{A}}$ is the Avogadro constant, $e$ the elementary charge, and $k_{\text{B}}$ the Boltzmann's constant. The $\mp$ sign in Eqn.~\ref{eq:el} are two different model assumptions, which match with separation distances larger than the double layer thickness~\cite{Wirix-Speetjens2005}. In Section~\ref{ssec:res_equilibrium} we show the difference and the significance for our simulation model. From an experimental view, surface potentials are not easily accessible. Therefore, $\Psi_{\text{p}}$ and $\Psi_{\text{s}}$ are typically approximated by the respective zeta potentials $\zeta_{\mathrm{p}}$ and $\zeta_{\mathrm{s}}$~\cite{Reginka2021}. The zeta potential describes the electric potential present at the shearing plane between mobile and immobile ions surrounding a fluid-immersed surface~\cite{Butt2003}. It can be experimentally quantified by light scattering of electrophoretically moved particles together with a fitting electrokinetic treatment~\cite{Delgado2007}.

\subsection{Magnetic particles\label{ssec:methods_magn}}
Superparamagnetic Dynabead M-270 Carboxylic Acid particles from Thermo Fisher with a diameter of \SI{2.8}{\micro\meter} in distilled water have been used. An effective magnetic field $\vec H_\mathrm{eff}$ induces a magnetic moment $\vec m$ inside the particle proportional to its volume $V$ and magnetic susceptibility $\chi$. 

\begin{equation}
\vec m = V \chi \vec H_\mathrm{eff}
\label{equ_susc}
\end{equation}

$\vec H_\mathrm{eff}$ results from the magnetic stray field landscape and the alternating applied external field  $\vec H_\mathrm{ext}$. Magnetic field contributions of neighboring particles will be neglected in this work. With small applied fields, the magnetic susceptibility can be assumed to be constant, but with increasing field, the particle's magnetic moment saturates. The resulting magnetization curve can be approximated by the Langevin function, shown for Dynabeads M-270 in~\cite{chen2015magnetic}, where both, the dimensionless parameter $a$ and the saturation moment $m_\mathrm{s}$ need to be tuned to fit the experimental measurements.

\begin{equation}
\vec m = m_\mathrm{s} \left(\coth\left(\frac{\vec H_\mathrm{eff}}{a}\right)-\frac{a}{\vec H_\mathrm{eff}}\right)
\label{equ_langevin}
\end{equation}

The magnetic gradient force $\vec F_\mathrm{m}$ on the particle is computed by the negative gradient of the energy of its magnetic moment $\vec m$ in the field $\mu_0\vec H_\mathrm{eff}$ with the permeability of free space $\mu_0$.

\begin{equation}
\vec F_\mathrm{m}=\mu_0\nabla(\vec m\vec H_\mathrm{eff})
\label{equ_Fm}
\end{equation}

Please keep in mind that we compute $\vec m$ and $\vec F_\mathrm{m}$ in the center of the superparamagnetic particle, assuming a homogeneous field and magnetization inside the particle. Dynabeads M-270 consist of iron oxide nanoparticles trapped inside the micron-sized particle by a polymer 
shell. The iron oxide nanoparticles can form small clusters~\cite{fonnum2005characterisation}, which may lead to an inhomogeneous magnetization inside the particle. In Section~\ref{ssec:res_MP_working_point}, we average multiple points on the surface of a magnetic particle to examine the influence of an inhomogeneous field on $\vec F_\mathrm{m}$.

\subsection{Lattice Boltzmann fluid dynamics\label{ssec:methods_LBM}}
The lattice Boltzmann method is based on a discretized lattice grid with uniform grid length. Fictive fluid molecules on the grid points perform consecutive propagation and collision. Macroscopic fluid properties can be obtained by distribution functions $f_k$ of these fictive molecules in space $x$ and time $t$. 

\begin{equation}
\underbrace{\vphantom{\frac{1}{\tau}}f_k( x+ e_k \delta_t,t+\delta_t) = f_k( x,t)}_{\mathrm{streaming}} - \underbrace{\frac{(f_k( x,t) - f_k^{\mathrm{eq}}( x,t))}{\tau}}_\mathrm{collision} + \underbrace{\vphantom{\frac{1}{\tau}}F_k( x,t)}_\mathrm{external\ forces}
\label{equation_lbm}
\end{equation} 

ESPResSo uses 3 dimensions with 19 discrete velocities, $k=0\dots 18$, the D3Q19 version of the lattice Boltzmann method~\cite{dunweg2009lattice}. $e_k$ is the discrete velocity vector, $\delta_t$ is the time step, $\tau$ denotes the relaxation time, $f_k^{\mathrm{eq}}$ is the equilibrium function depending on macroscopic variables velocity $v(t)$ and density $\rho (t)$.

Magnetic particles are 2-way coupled with the lattice Boltzmann fluid. The external force term $F_k(x,t)$ is the sum of all direct, magnetic, and DLVO forces acting on the particle. The fictive fluid molecules are propagated accordingly and the fluid velocity $v_\mathrm{f}(x,t)$ of the lattice grid points is updated. $F_k(x,t)$ and $v(x,t)$ are linearly interpolated in both directions, $F_k(x,t)$ from the moving Lagrangian center of particle to lattice grid points, and $v_\mathrm{f}(x,t)$ vice versa. The interpolated fluid velocity at the particles center position $\vec v_\mathrm{f}$ is used to calculate the fluidic drag force $\vec F_\mathrm{d}$ on the particle.

\begin{equation}
\vec F_\mathrm{d} = -\gamma (\vec v_\mathrm{p} - \vec v_\mathrm{f})
\label{equ_Fd}
\end{equation}

$\vec F_\mathrm{d}$ is inversely proportional to the velocity difference of the particle velocity $\vec v_\mathrm{p}$ and the fluid velocity $\vec v_\mathrm{f}$ at the specific particle position multiplied by the drag coefficient $\gamma$, which is given in Ns/m units~\cite{cimrak2012modelling}. Due to the discretization of the fluid lattice and the representation of the magnetic particle by mass points, $\gamma$ needs to be calibrated, which is described in Section~\ref{ssec:res_calibration_drag}. 


Please note, that ESPResSo does not use predefined units. These can be set according to the underlying simulated phenomenon. In this work we set the time scale to \SI{0.1}{\micro\second}, length scale to \SI{0.1}{\micro\meter} and mass scale to \SI{e-15}{\kilogram}. Other physical quantities are either not scaled or a proper combination of time, length, and mass units.

The fluid parameters in this work are set for water at \SI{20}{\degreeCelsius} with density $\rho_\mathrm{F}~=~\SI{1000}{\kilogram\per\meter\cubed}$, dynamic viscosity $\mu~=~\SI{0.001}{\kilogram\per\meter\per\second}$ and relative permittivity $\epsilon_\mathrm{r}~=~80.1$.  

\subsection{Magnetic stray field landscape\label{ssec:methods_MFL}}
In this work, we use an analytic solution to obtain the magnetic stray field landscape generated by a parallel-stripe domain pattern with remanent in-plane magnetizations forming head-to-head and tail-to-tail domain walls with flat interfaces. 

In micromagnetic simulations, we are limited in simulation size~\cite{gusenbauer2020extracting}. The dimension of finite elements to resolve nucleation and domain wall movement is typically a few nm only. To consider the size of the original bilayer system, we are using an analytic solution of the magnetic field created by cuboidal permanent magnets~\cite{akoun19843d}. 

The magnetization inside these magnets is assumed to be homogeneous and parallel to one of the cuboids axes, in our case the $x$-axis (compare Fig.~\ref{fig:model}). Since the magnetic polarization $\vec J$ is perpendicular to the interfaces between the stripes, these faces have a surface charge density $\sigma=|\vec J|$. The magnetic field by uniformly charged rectangular surfaces can be calculated analytically at each point in space, which is shown by~\cite{akoun19843d}. Superposing all charged surface fields, the full stray field landscape is obtained. 

With this, it is possible to treat particle motion in a magnetic stray field landscape created by the domain configuration of Fig.~\ref{fig:model} with a depth along the $y$-coordinate of \SI{50}{\micro\meter}. 
Note that two touching domains with opposite magnetic orientation form a domain wall. By changing the magnetic configuration, e.g., applying varying external magnetic fields, this domain wall can move. Domain wall formation is not considered when treating the magnetic stray field landscape by analytic equations. Therefore, we talk about interfaces between adjacent stripe domains in the simulations. These interfaces are treated as infinitely thin.

Here, we simulate a single jump from a head-to-head to a tail-to-tail interface above the bilayer system. The magnetic field landscape considers the superposition of the related interfaces and all other interfaces in the system. Including domain wall formation would reduce the magnitude of the resulting stray field above an interface. In this work, this reduction is accounted for by an additional thickness of the polymer spacer layer.



\subsection{Minor force contributions\label{ssec:methods_minor}}
In this section, we shortly summarize additional force contributions acting on the magnetic particles. 

Immersing a particle into a fluid, without applying external force terms like DLVO and magnetic forces, results in a final steady-state velocity based on buoyancy and gravitation. The steady-state velocity $v_\mathrm{s}$ is defined as~\cite{SOUSA2020215}

\begin{equation}
v_\mathrm{s} = \frac{d^2(\rho_\mathrm{P} - \rho_\mathrm{F})g}{18\mu}
\label{equ_Fbg}
\end{equation}

with the diameter $d$, the density difference of particle and fluid, $\rho_\mathrm{P}-\rho_\mathrm{F}$, the gravitational acceleration $g$ and the fluid viscosity $\mu$. With the standard parameters defined in Table~\ref{tab:initial_parameters} we obtain $v_\mathrm{s}=\SI{2.56}{\micro\meter\per\second}$, which is 2 orders of magnitude smaller than the obtained velocities in the particle trajectory experiments (compare also Section~\ref{ssec:res_calibration_drag} and the resulting force contributions in Section~\ref{ssec:res_equilibrium}).

As we consider only a single particle (accounting for a large distance between individual particles), we neglect possible field contributions, electrostatic potentials, or adhesion/repelling of neighboring particles. In case of a more dense particle suspension, we refer to preliminary studies, for instance, in~\cite{gusenbauer2018simulation} and~\cite{huhnstock2024combined}.

Brownian motion, the random movement of particles due to thermal fluctuations, is not treated in this work. Fonnum and co-workers suggest treating the stochastic Brownian motion up to a particle diameter of \SI{1}{\micro\meter}~\cite{fonnum2005characterisation}, because the effect becomes negligible with increasing particle size, which is the case with our used particles.

\section{\label{sec:results_calibration}Model calibration}

The simulation environment used for the present calculations requires several calibration steps, which will be described in detail in this section.


\subsection{Calibration of fluid-particle interaction\label{ssec:res_calibration_drag}}
In the simulation not only the motion of the particle against quiescent fluid volume-elements far away from the particle is considered, but also the back-action of a moving particle to the velocity of close-by surrounding fluid volume elements. The velocity differences between particles and different fluid volume-elements, therefore, vary with their positions and discretization. To describe the drag coefficient of Eqn.~\ref{equ_Fd} correctly using one coefficient (but accounting for the different relative velocities), this coefficient needs to be carefully calibrated using reasonable assumptions. Similar to ~\cite{gusenbauer2018simulation}, we performed pushing simulations with an initial velocity $v_0$ and compared the deceleration with an analytical solution of the equation of motion. In the present work, in addition, we accelerate the particle by a constant force and compare the particle's trajectory with the analytical solution of the corresponding equation of motion.

For a spherical particle, the classical fluidic drag force is calculated by 
\begin{align}
\vec F_\mathrm{d} = - 6 \pi \mu r \vec v
\label{eqn:stokes}
\end{align}

with the particle radius $r$, the dynamic viscosity of the fluid $\mu$ and the relative velocity $\vec v$ of the particle with respect to the fluid. We performed the tests in static fluid and only in a single direction, hence the scalar velocity $v$ is equal to the particle velocity 
and we apply an initial velocity $v_0~=~\SI{200}{\micro\meter\per\second}$ for decelerating, and $v_0~=~\SI{0}{\micro\meter\per\second}$ and a steady-state velocity (at equilibrium of driving and drag forces) $v_\mathrm{s}~=~\SI{200}{\micro\meter\per\second}$ for accelerating the particle. The chosen velocities are in the same range as the ones observed in the particle transport experiments. The analytical solutions are obtained from 
differential equations with the basic equation of motion of a particle with mass $m$, for deceleration ($0 = m \dot{v} + 6 \pi \mu r v$) and acceleration ($6 \pi \mu r v_\mathrm{s} = m \dot{v} + 6 \pi \mu r v$), respectively. The solution of both equations 
are simple exponential equations 

\begin{align}
v_\mathrm{dec} = v_0 \exp\left(-\frac{6\pi\mu r}{m} t\right)
\label{eqn_deceleration}
\end{align}


\begin{align}
v_\mathrm{acc} = v_s \left(1-\exp\left(-\frac{6\pi\mu r}{m} t\right)\right)
\label{eqn_acceleration}
\end{align}

\begin{figure}
    \centering
    \includegraphics[width=0.45\textwidth]{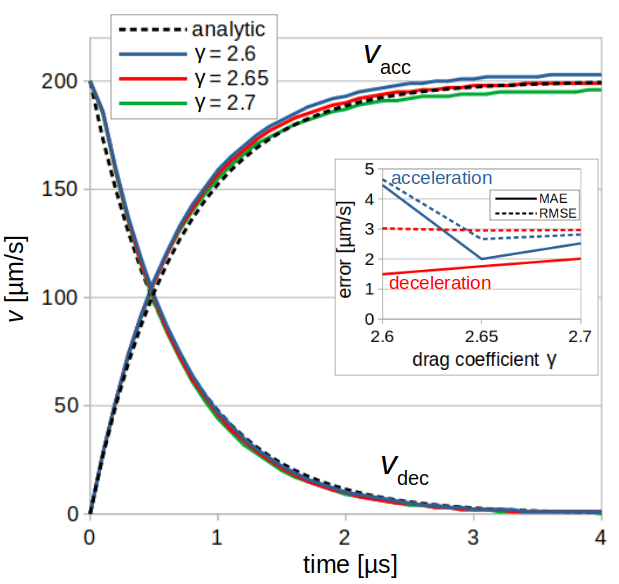}
    \caption{Calibration of the fluidic drag coefficient $\gamma$ in Eqn.~\ref{equ_Fd} is shown for deceleration ($v_{\mathrm{dec}}$) and acceleration ($v_{\mathrm{acc}}$) curves of a particle. Different $\gamma$ values are compared to the exact analytical solutions. Mean absolute error (MAE) and root mean square error (RMSE) are shown for both curves in the inset.}
    \label{fig:drag_calibration}
\end{figure}

In Fig.~\ref{fig:drag_calibration}, we compare the analytical solutions with simulations to calibrate the drag coefficient $\gamma$ of Eqn.~\ref{equ_Fd}. For the simulations we place a particle in the center of a simulation box and apply $v_0~=~\SI{200}{\micro\meter\per\second}$ for deceleration and a constant force $\vec F_\mathrm{d}$ with $v_\mathrm{s}~=~\SI{200}{\micro\meter\per\second}$ for acceleration. 
In the inset of Fig.~\ref{fig:drag_calibration}, we show the mean absolute error (MAE) and the root mean square error (RMSE), calculated for \SI{4}{\micro\second} in 40 time steps, to find the best fit of the simulated curves to the analytical ones.

Given that the deceleration simulations appear mostly insensitive to $\gamma$, the acceleration fit provides a more meaningful setup for fitting the parameter. Also, in the acceleration curves, a clear difference in settling velocity $v_\mathrm{s}$ can be observed compared to the analytic solution with deviations of about $\pm$\SI{3}{\micro\meter\per\second} for $\gamma = 2.6$ Ns/m and $\gamma = 2.7$ Ns/m, respectively.


The discretization of the fluid lattice needs to be chosen such that the particle diameter is smaller than the lattice grid length, which means a single particle should fit multiple times into a single fluid lattice cell~\cite{gusenbauer2018simulation}. We decided to use a lattice grid length of \SI{10}{\micro\meter}, which has a good ratio to the particle diameter with \SI{2.8}{\micro\meter}, and a simulation box of 50$\times$50$\times$100~$\upmu$m$^3$. The calibrated $\gamma$ value was found to be 2.65 Ns/m. Please note that the calibrated drag coefficient depends on the discretization of the lattice and needs to be recalibrated when the grid length is changed.

\subsection{Induced magnetic moment of Dynabeads M-270\label{ssec:res_MP_working_point}}
\begin{figure}
    \centering
    \includegraphics[width=0.35\textwidth]{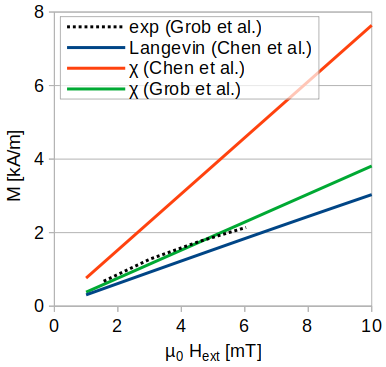}
    \caption{Different approximations of initial magnetization curves of Dynabead M-270. Experimental measurement of Grob et al.~\cite{grob2018magnetic} is compared to the constant magnetic susceptibility $\chi$ representation and the Langevin function of Chen et al.~\cite{chen2015magnetic}.}
    \label{fig:particle_magnetization}
\end{figure}

Now, we need to calibrate the magnetization properties of the used superparamagnetic particle.
A particle of the present experiment is subjected to an effective magnetic field of a few mT. The induced magnetization of the particle in this field range is shown in Fig.~\ref{fig:particle_magnetization}. The slope of the magnetization curve is defined by the magnetic susceptibility $\chi$. Small variations in the given measurements, especially with small applied magnetic fields, can be observed, which affect the correct particle trajectory in the simulations. 

In this work, we apply external magnetic fields between $\pm$\SI{8}{\milli\tesla}. Grob and co-workers show a magnetization curve of Dynabeads M-270 between $\pm$\SI{6}{\milli\tesla}~\cite{grob2018magnetic}, which can be approximated with a constant magnetic susceptibility $\chi=0.479$, and a given saturation magnetization $M_\mathrm{s}=\SI{10.24}{\kilo\ampere\per\meter}$. They measured the properties with a Quantum Design MPMS-XL SQUID magnetometer at room temperature. In the work of Chen et al. they quote the vendor's magnetic properties~\cite{chen2015magnetic}, which are $\chi~=~0.96$ and $M_\mathrm{s}~=~\SI{20.8}{\kilo\ampere\per\meter}$. They also give the approximated parameters for the Langevin function (Eqn.~\ref{equ_langevin}), which are $\upmu_0 a~=~0.021$ and $m_\mathrm{s}~=~\SI{0.223}{\pico\ampere\meter\squared}$. Note that the properties for $\chi$ and $M_\mathrm{s}$ differ by about a factor of 2 between the two groups. 

In Fig.~\ref{fig:particle_magnetization}, we compare the different approximations to the measured magnetization curve up to \SI{10}{\milli\tesla}. A much steeper curve is shown in \cite{chen2015magnetic} with constant $\chi$. Even though the magnetic saturation is lower in \cite{grob2018magnetic}, the Langevin approximation of \cite{chen2015magnetic} is lower at these small applied fields but overtake with larger fields to reach saturation, which is not shown in the plot. We assume that the value of $\chi$ given in \cite{chen2015magnetic} is overestimated in the field range we are working on, or in other words, a linear fit was taken at much lower fields. A linear fit of the Langevin function in the given field range results in $\chi=0.383$, which is much closer to the experimental measurements. In the results section, we demonstrate the effect of varying $\chi$ and $M_\mathrm{s}$ on the resulting particle trajectories.

With the point-dipole approximation of the magnetic particle, the stray field information is taken once at the center, from which the magnetization of the particle is calculated. To evaluate the effect of stray field discretization and inhomogeneous magnetization inside the particle we evaluated the stray field multiple times at the particle's surface and averaged the magnetization in the center of the particle. Averaging fields at multiple points of the surface has the advantage that the discretization of the stray field is smoothed, but it increases simulation time. Particle trajectories of both scenarios were congruent, so we use the point-dipole approximation in the rest of this work. 


\section{\label{sec:results_trajectory}Comparison of experimental and simulated particle trajectories}

In this section, we demonstrate the effect of varying the model parameters on a particle's equilibrium height above the polymer surface when resting, and the jump height as well as jump duration initialized by switching the $z$-component of the external magnetic field. 

\subsection{Experimentally determined particle jump}\label{ssec:res_exp_jump}

\begin{figure}[h!]
    \centering
    \includegraphics[width=\linewidth]{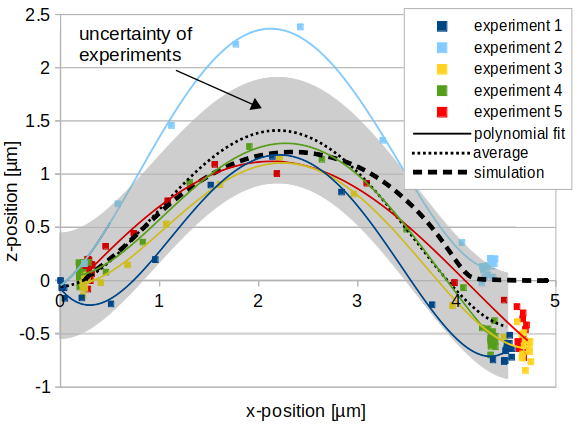}
    \caption{Vertical trajectories of Dynabead M-270 superparamagnetic particles moving on top of a polymer-covered, magnetic stripe domain patterned substrate after applying time-varying external magnetic fields. The evolution of the vertical $z$-position as a function of lateral $x$-position was determined by optical defocusing-based 3D particle tracking. Measured data points (squares) and a 4th-degree polynomial fit (solid lines) are compared to a simulation with standard parameters (thick dashed line) shown in Tab.~\ref{tab:initial_parameters}. The average of all polynomial fits is given as a dotted line, with the shaded area representing the experimental uncertainty. The $z$-position was determined for the center of the particles and normalized to zero. The $x$-position is given from one interface (domain wall) to the next one within the magnetically patterned substrate. }
    \label{fig:exp_result}
\end{figure}

The motion of water-dispersed superparamagnetic particles on top of the magnetically patterned substrate was induced by magnetic field pulses in $z$- and $x$-directions of $\upmu_0 H_{\mathrm{z,max}}$~=~\SI{8}{\milli\tesla} and $\upmu_0 H_{\mathrm{x,max}}$~=~\SI{1}{\milli\tesla}, respectively. Each field pulse in the $z$-direction forced a particle jump between positions of neighboring domain walls (denoted as interfaces in the simulation), e.g., from a head-to-head wall to a tail-to-tail wall. With a positive $x$-component of the field, a forward jump in the +$x$-direction can be chosen. We determined the jump trajectory for such a motion event by applying an optical defocusing-based 3D particle tracking method, in analogy to the results presented in \cite{huhnstock2022three}. 

The $z$-coordinates of 5 individual particles measured for different locations on the substrate are shown in Fig.~\ref{fig:exp_result} as a function of the lateral $x$ position. The jump height and duration were determined for each particle and subsequently averaged to obtain about \SI{1.4}{\micro\meter} as the average jump height and about \SI{7}{\milli\second} as the average jump duration. Differences in individual particle trajectories arise from variations in the magnetic properties of the particles (these particles typically exhibit a distribution in size and magnetic content) and variations in the magnetic stray field landscape caused by different domain wall properties, which is inherent to the fabrication procedure. By averaging over different particles and domain wall positions, we try to account for these variations. 
The final $z$-positions after the jump show a slight deviation as compared to the start $z$-positions before the jump, i.e., the particles end up at lower $z$-positions compared to when they started. The defocusing-based 3D tracking method is sensitive to the lighting conditions in the experiment, thus, we mainly explain this deviation with inhomogeneous sample illumination. It has no significant impact on the determination of jump duration; for the jump height, it is accounted for by the experimental uncertainty, shown as the shaded area in Fig.~\ref{fig:exp_result} for the averaged experimental data.

For comparison, we show the result of a simulation with standard model parameters (dashed line), which we will introduce and discuss in the next section.
Note that due to the experimental measurement uncertainty, we cannot resolve changes in particle $z$-position that are lower than \SI{500}{\nano\meter}. Particle jumps with a height below this uncertainty are, therefore, solely measured as a lateral movement in the $x$-direction.

\subsection{Simulation of a particle jump\label{ssec:res_equilibrium}}
\begin{figure}[ht]
    \centering
    \includegraphics[width=\linewidth]{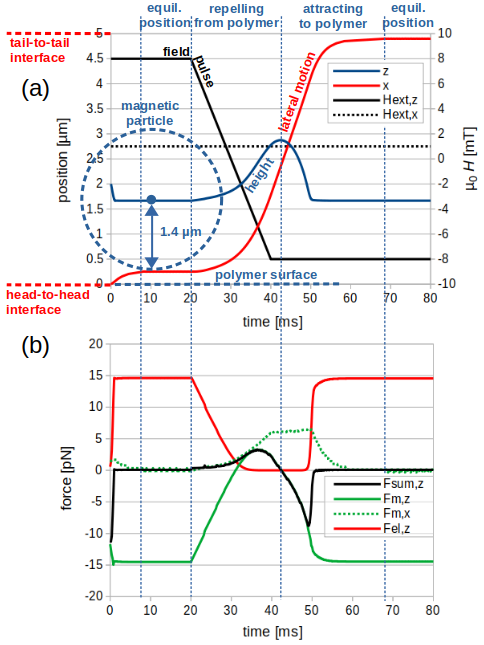}
    \caption{Simulation result with standard parameter set. (a) A magnetic particle's trajectory ($z$- and $x$-position) together with the applied external magnetic fields ($H_{\mathrm{ext,z}}$ and $H_{\mathrm{ext,x}}$), and (b) force contributions are shown for the duration of a single particle jump. A blue dashed circle at equilibrium $z$-position in (a) depicts the particle size and demonstrates the gap between the particle surface and the polymer surface. Interface positions of touching stripes (horizontal red dashed lines) are important for the particle's lateral $x$-position. Vertical dashed lines show the different temporal force regimes from equilibrium position, repelling from the polymer surface, attracting to the polymer surface to the next equilibrium position. The Van-der-Waals force $\vec F_\mathrm{vdW}$ and the force from gravity and buoyancy $\vec F_\mathrm{gb}$ are negligible and not shown. Note, that $\vec F_\mathrm{el}$ is used with the negative sign of Eqn.~\ref{eq:el}.}
    \label{fig:start_simulation}
\end{figure}
We define a set of initial parameters in Tab.~\ref{tab:initial_parameters} as a starting point for our theoretical study. The computed particle trajectory is compared to the experimental result. By varying the simulation parameters, we examine their impact and assess the model's sensitivity to these changes.

\begin{table*}[h]
    \centering
    \begin{tabular}{lcrll}
        \bf{Parameter} &  & \bf{Value} & \bf{Unit} & \bf{Comment}\\
         \hline
        \bf{ESPResSo unit conversion} &  &  & &\\
        length & $l$ & $10^{-7}$ &   & $\hat{=}$ 0.1 $\upmu$m\\
        time & $t$ & $10^{-7}$ &   & $\hat{=}$ 0.1 $\upmu$s\\
        weight & $w$ & $10^{-15}$ &   & $\hat{=}$ 10$^{-15}$ kg\\
        \hline
        \bf{Fluid} &  &  &&\\
        channel & & 50x50x100 & $\upmu$m$^3$ & \\
         grid length &  & 100 & &  $\hat{=}$ 10 $\upmu$m lattice grid length\\
         drag coefficient & $\gamma$ & 2.65  & Ns/m & see calibration in Section.~\ref{ssec:res_calibration_drag}\\
        *density & $\rho_\mathrm{F}$ & 1000 & kg/m³ &water 20$^\circ$C\\
        *dynamic viscosity & $\mu$ & 0.001 & kg/(ms) &water 20$^\circ$C\\
        *relative permittivity & $\epsilon_\mathrm{r}$ & 80.1  &&water 20$^\circ$C\\
        \hline
        \bf{Dynabeads M-270} & &  & & other particles may lead to different trajectories\\
         diameter & $d$ & 2.8 & $\upmu$m  &\\
         density & $\rho_\mathrm{P}$ & 1600 & kg/m$^3$ &\\
         saturation magnetization & $M_\mathrm{s}$ & 10.24 & kA/m  &\cite{grob2018magnetic}, compare Section~\ref{ssec:res_MP_working_point}\\
         magnetic susceptibility & $\chi$ & 0.479  & & \cite{grob2018magnetic}, compare Section~\ref{ssec:res_MP_working_point}\\ 
         \hline
         \bf{Magnetic field} & &  & &\\
         *external magnetic field $z$ & $\upmu_0 H_\mathrm{ext,z}$ & $\pm8$& mT & from positive to negative field in 100 steps\\
         *external magnetic field $x$ & $\upmu_0 H_\mathrm{ext,x}$ & 1 & mT & \\
         *switching time & $t_\mathrm{s}$ & 20 & ms & time for $\upmu_0 H_\mathrm{ext,z}$ to change the sign\\
        \hline
        \bf{DLVO} &  &   &&\\
        Hamaker constant & $A_\mathrm{132}$ & 1.23e-20 & J  &\cite{huhnstock2022three}, $A_\mathrm{132}$ depends on fluid, particle and substrate properties \\
        *Debye length & $1/\kappa$ & 100 & nm  & \\
        *Particle zeta potential & $\zeta_\mathrm{P}$ & -28.28 & mV &\cite{wise2015magnetophoretic}\\
        *Substrate zeta potential & $\zeta_\mathrm{S}$ & -65 & mV &\cite{huhnstock2022three}\\
        *polymer thickness & t$_\mathrm{PMMA}$ & 2 & $\upmu$m & spacer layer between magnetic bilayer and fluid channel,\\
        &&&& artificially increased due to the overestimated analytic stray field
    \end{tabular}
    \caption{Initial parameter set of the simulation model. Parameters marked with * can be tuned in the experiment.}
    \label{tab:initial_parameters}
\end{table*}

In Fig.~\ref{fig:start_simulation}(a), we show the temporal evolution of a jump simulation for the parameters of Tab.~\ref{tab:initial_parameters}. They have been chosen according to experimental measurements (compare Section~\ref{ssec:methods_experiment} and~\ref{ssec:res_exp_jump}), literature investigation, calibration of the simulation model, and by selecting an accurate working point of the magnetic particle (compare Section~\ref{sec:results_calibration}). All parameters in Tab.~\ref{tab:initial_parameters} marked with a * can be tuned in the experiment but are not necessarily easy to determine.

We start with an initial particle position right above a head-to-head interface at a distance of \SI{2}{\micro\meter} to the polymer layer and a positive external magnetic field $\upmu_0 H_\mathrm{ext,z}$ = \SI{8}{\milli\tesla} and $\upmu_0 H_\mathrm{ext,x}$ = \SI{1}{\milli\tesla}. First the particle relaxes to find its equilibrium position and than we apply a single field pulse reversing $H_\mathrm{ext,z}$ from +\SI{8}{\milli\tesla} to -\SI{8}{\milli\tesla} within \SI{20}{\milli\second} at constant $H_\mathrm{ext,x}$. Note that the particle position is given for the center of the particle. The magnetization of the particle is pointing in the direction of the external magnetic field (compare Section~\ref{ssec:res_MP_working_point}). 

With positive field values and a positive magnetization, the particle is attracted to a head-to-head interface. After only a few ms, the particle finds an equilibrium $x$- and $z$-position, which is slightly shifted to the positive $x$-direction away from the head-to-head interface and with a certain distance $z$ above the polymer surface. With the parameters of Tab.~\ref{tab:initial_parameters}, the equilibrium distance between particle surface and polymer surface is about \SI{260}{\nano\meter} (compare blue dashed lines for particle and polymer surface in Fig.~\ref{fig:start_simulation}(a)). The equilibrium height is defined by balancing the magnetic force $\vec F_\mathrm{m}$ and the electrostatic force $\vec F_\mathrm{el}$ (Fig.~\ref{fig:start_simulation}(b)), compare also Fig.~\ref{fig:force_distance}). We are using $\vec F_\mathrm{el}$ with the negative sign of Eqn.~\ref{eq:el}. Using the positive sign is resulting in a slightly different equilibrium height only. The rest of the particle trajectory is about the same. All other force contributions, the van-der-Waals force $\vec F_\mathrm{vdW}$ and the force from gravity and buoyancy $\vec F_\mathrm{gb}$ are negligible small. A comparison of all force contributions with given distance between polymer and particle surface is shown and discussed in Section~\ref{ssec:res_uncertainties}.

The particle starts to move from head-to-head to the next tail-to-tail interface, whereby the jump direction is controlled by $H_\mathrm{ext,x}$. The particle height is slowly increasing with a force balance of $\vec F_\mathrm{el}$ and $\vec F_\mathrm{m}$ (Fig.~\ref{fig:start_simulation}(b)). This initial increase of particle height cannot be seen in the experiments due to the measurement uncertainty. When the external field $H_\mathrm{ext,z}$ is pointing in the negative direction, the particle magnetization is reversed as well, and the particle is repelled by the head-to-head interface. 
$\vec F_\mathrm{m}$ gets dominant (Fig.~\ref{fig:start_simulation}(b)) and the particle height is increasing faster. The particle reaches a maximum height of about \SI{1.5}{\micro\meter} above the equilibrium height. The maximum height is reached between the head-to-head and tail-to-tail interfaces when the repulsion of the head-to-head interface is overcome by the attraction of the tail-to-tail interface. $\vec F_\mathrm{el}$ becomes more influential, and the equilibrium position is found close to the tail-to-tail interface. 
Please note that we do not show the fluid force contribution $\vec F_\mathrm{d}$, as it is handled with the 2-way coupling of the lattice Boltzmann method described in Section~\ref{ssec:methods_LBM}.

In the supplementary materials, we show a video of a full simulation run with the standard parameters shown in Tab.~\ref{tab:initial_parameters}. 





\subsection{Variation of model parameters \label{ssec:res_variation}}

\begin{figure*}[ht]
    \centering
    \includegraphics[width=\textwidth]{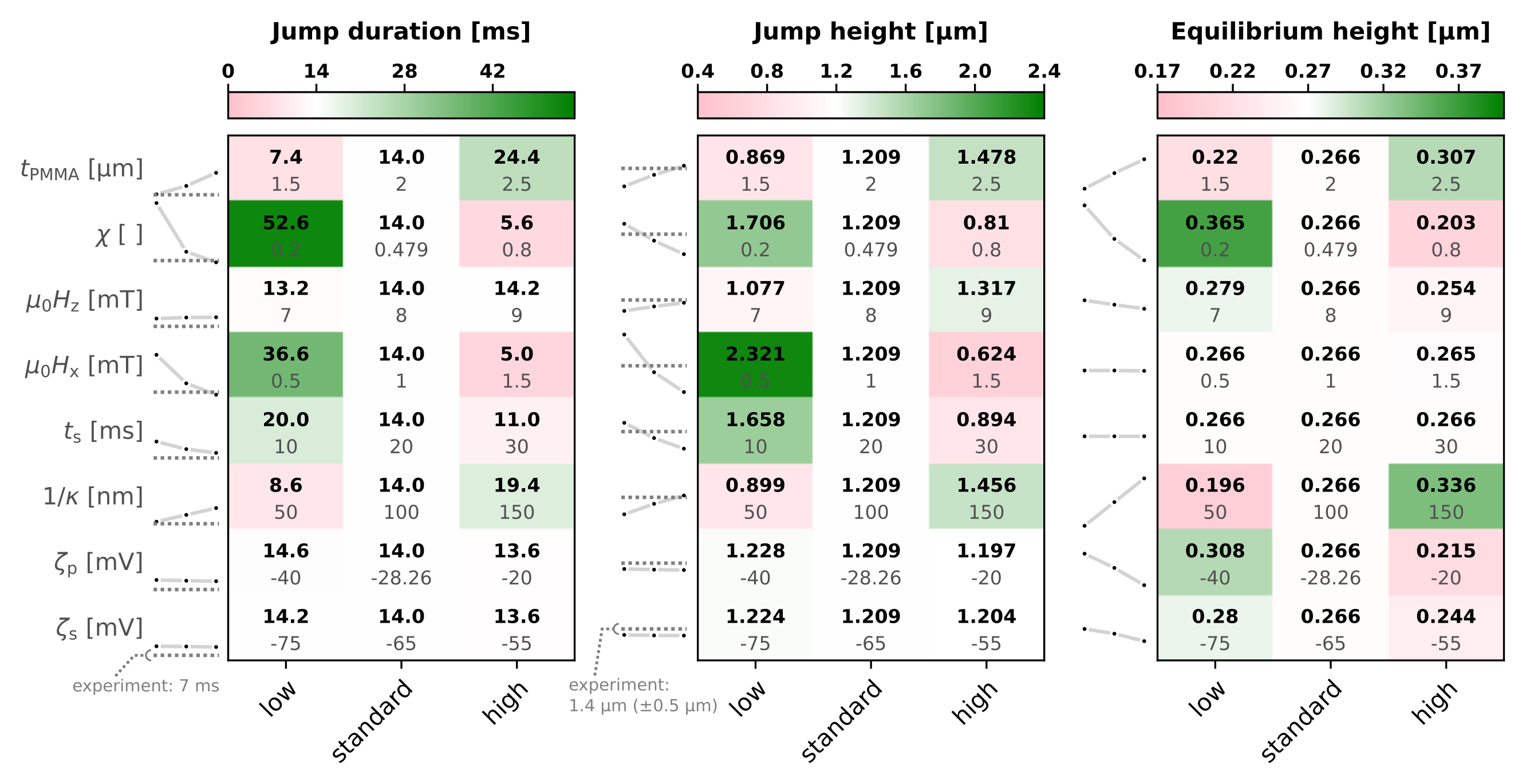}
    \caption{Effect of parameter variation on the characteristic quantities jump duration, jump height and equilibrium height for a simulated particle jump. The bold numbers and the color code represent the respective quantity. The gray numbers below the simulation results give the low, standard and high values of the specific parameter on the left. Here,  $t_{\mathrm{PMMA}}$ is the polymer layer thickness varied between \SI{1.5}{\micro\meter} to \SI{2.5}{\micro\meter}, $\chi$ is the particle's magnetic susceptibility, varied between 0.2 and 0.8, $z$- and $x$- components of the external field have been varied between \SI{7}{\milli\tesla} and \SI{9}{\milli\tesla} ($z$-component) and \SI{0.5}{\milli\tesla} and \SI{1.5}{\milli\tesla} ($x$-component), the switching time $t_{\mathrm{s}}$ for the external field has been varied between \SI{10}{\milli\second} and \SI{30}{\milli\second}, the Debye length 1/$\kappa$ for the electrostatic interaction has been varied between \SI{50}{\nano\meter} and \SI{150}{\nano\meter}, and the zeta potentials for particle $\zeta_{\mathrm{P}}$ and substrate $\zeta_{\mathrm{S}}$ surfaces have been varied between \SI{-40}{\milli\volt} and \SI{-20}{\milli\volt} (particle) and \SI{-75}{\milli\volt} and \SI{-55}{\milli\volt} (substrate). A trendline of the results is compared to the experimental values shown as a dashed line. Note that the equilibrium height cannot be experimentally measured.}   
    \label{fig:parameter_variation}
\end{figure*}
In the previous sections, we have shown that the simulation model for predicting magnetic particle trajectories describes the experiments qualitatively well. It contains, however, a larger number of parameters, rendering a quantitative comparison challenging. We, therefore, investigate the sensitivity of the simulated jump height, duration, and particle equilibrium height before and after the jump on these parameters. Fig.~\ref{fig:parameter_variation} shows the variation of the individual parameters, keeping all other parameters at their standard values (Tab.~\ref{tab:initial_parameters}). The simulation results for jump duration and height are compared to the experiments, which is not possible for the equilibrium height, as we could not experimentally quantify this property so far. When determining the jump duration from the simulated trajectories, we account for the experimental $z$-position measurement uncertainty of \SI{500}{\nano\meter} to perform a sensible comparison. Hence, we count the start of the simulated particle jump as soon as its $z$-position increase by more than \SI{500}{\nano\meter} after reversing $H_\mathrm{ext,z}$.

\subsubsection{Polymer layer thickness $t_{\mathrm{PMMA}}$}\label{sssec:pmmathickness}
An increase of the gap between the ferromagnetic layer of the underlying thin film system and the particle, realized by an increase of the polymer thickness from \SI{1.5}{\micro\meter} to \SI{2.5}{\micro\meter}, reduces the magnetic stray field while maintaining the repulsive electrostatic particle surface-polymer surface interaction. This pushes the particle to a higher equilibrium position and induces a slightly higher and longer jump. 

\subsubsection{Magnetic susceptibility $\chi$}\label{sssec:susceptibility}
A reduction in the magnetic susceptibility of the particle from 0.8 to 0.2 drastically increases the resulting characteristics. The particle's magnetic moment is reduced, leading to a decreased magnetic force at maintained electrostatic repulsion from the polymer surface and, therefore, a higher equilibrium position, a larger jump height, and a remarkably increased jump duration.

\subsubsection{External fields $H_\mathrm{ext,z}$ and $H_\mathrm{ext,x}$}\label{sssec:externalfields}
The $z$-component of the external field has negligible effects in the range between \SI{9}{\milli\tesla} and \SI{7}{\milli\tesla}, while the $x$-component seems to be of major importance. Only a small variation of $H_\mathrm{ext,x}$ yielded significant changes for the jump duration and height. $H_\mathrm{ext,x}$ is necessary to control the path of the particle and causes a slight shift of the equilibrium $x$-position away from the interface center.
With $H_\mathrm{ext,x}=$ \SI{0.5}{\milli\tesla}, the equilibrium $x$-position is closer to the interface, which results in a higher $z$-component and a smaller $x$-component of the magnetic gradient field during the main jump. Consequently, the jump is higher but lasts longer. With $H_\mathrm{ext,x}=$ \SI{1.5}{\milli\tesla} the effect is reversed.

\subsubsection{Switching time $t_{\mathrm{s}}$}\label{sssec:switchingtime}
The jump height depends sensitively on the switching time of the external $z$-field between the two plateaus $H_{\mathrm{z,max}}$ and $-H_{\mathrm{z,max}}$. With a faster switch (\SI{10}{\milli\second} switching time), the jump is much higher, whereas increasing the switching time up to \SI{30}{\milli\second} lowers the height. The jump duration and equilibrium position are about the same for the different switching times, whereas it is important to note that with larger switching times, the jump duration is slightly reduced.

\subsubsection{Debye length 1/$\kappa$ and zeta potentials $\zeta$}\label{sssec:debyelengthzetapotentials}
Increasing the Debye length from \SI{50}{\nano\meter} to \SI{150}{\nano\meter} increases all characteristic quantities likely due to a stronger electrostatic repulsion from the polymer surface, weakening the influence of the magnetic force. Lowering the magnitude of the zeta potential for the particle surface $\zeta_\mathrm{P}$ from \SI{-40}{\milli\volt} to \SI{-20}{\milli\volt} results in a small change of the equilibrium position due to the varied electrostatic force from \SI{308}{\nano\meter} to \SI{215}{\nano\meter}, respectively. The same trend is observable for changing the zeta potential of the substrate surface $\zeta_\mathrm{S}$ between \SI{-75}{\milli\volt} and \SI{-55}{\milli\volt}, whereas the changes in the equilibrium position are less significant in this case. The simulations yielded no significant changes in the jump height and duration when modifying the zeta potentials of the particle or the substrate surface.

\subsubsection{Other parameters}\label{sssec:otherparameters}
In the following, we discuss the influence of other parameters that are not shown in Fig.~\ref{fig:parameter_variation}. A variation of the particle's saturation magnetization $M_\mathrm{s}$ is negligible because the working point of the magnetic particle is in the increasing slope of the magnetization curve (compare Fig.~\ref{fig:particle_magnetization}). 


Changing the sign in Eqn.~\ref{eq:el} from negative to positive does not decisively alter the particle trajectory, but only slightly increases the equilibrium height by a few nm.

Results for a changing van-der-Waals force $\vec F_{\text{vdW}}$ are not shown, as it is orders of magnitude smaller than the other forces, but might have a major effect at very small equilibrium heights.

Gravitational and buoyancy effects are neglected as well, for similar reasons. 

Fluid properties and the calibration of the drag coefficient are discussed in Section~\ref{ssec:res_calibration_drag}. Changing one of the fluid parameters requires recalibration of the drag coefficient $\gamma$, which is not shown. A slight increase of $\gamma$ minorly raises the particle trajectory and vice versa. Particle properties, like diameter and density are kept constant as well to keep the drag calibration comprehensible.

\subsection{Uncertainties in the dynamics of magnetic particle close-to-substrate motion\label{ssec:res_uncertainties}}

This work demonstrates the effects of parameter variations on a simulated magnetic particle trajectory. Due to the lack of literature or experimental references, we discuss in the following sections several imponderables in the simulation model. 
For instance, the absolute height of a particle above the polymer layer is not known in the experiment, yet minor force contributions can be of major importance when a particle approaches the surface. Possible adherence of the particles on the polymer layer, permanent or temporary, affects the jump as well. Waiting for a specific equilibrium time is important, otherwise, jump trajectories are not reproducible due to different initial positions and magnetic gradient fields. 
\subsubsection{Reducing the equilibrium height}\label{sssec:reducedeqheight}
\begin{figure}[h!]
    \centering
    \includegraphics[width=\linewidth]{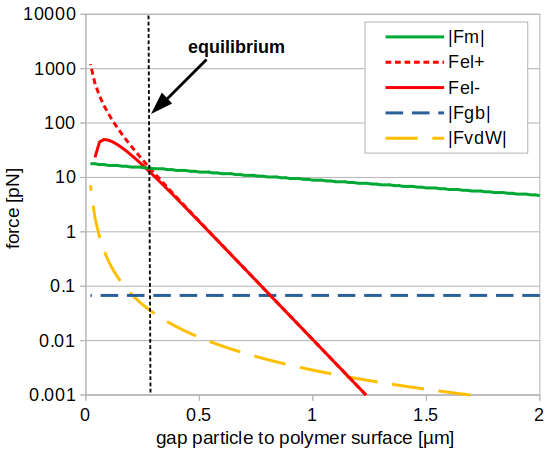}
    \caption{Force contributions in $z$-direction acting on a particle with a given gap between the particle surface and polymer surface. The sign of $\vec F_\mathrm{m}$, $\vec F_\mathrm{gb}$ and $\vec F_\mathrm{vdW}$ is negative, thus attracting the particle to the polymer surface. The equilibrium position for a particle is found with the sum of all forces (black dotted line).}
    \label{fig:force_distance}
\end{figure}
In case of a very short distance between the particle surface and polymer surface, more force terms are actively contributing to the particle trajectory. For the standard parameters used in this work, we can neglect the van-der-Waals force $\vec F_{\text{vdW}}$ and $\vec F_{\text{gb}}$ resulting from gravitation and buoyancy. The equilibrium position of a particle is then defined by the electrostatic and magnetic gradient forces.

For analysis of the force contributions, we place a particle inside the ESPResSo model next to a head-to-head interface. At the initial simulation time step we determine all contributing force terms. We start from contact with the polymer surface and repeat the analysis with increasing height up to \SI{2}{\micro\meter} in $z$-direction.
In Fig.~\ref{fig:force_distance}, the various force contributions are plotted against the changing gap between the particle surface and polymer surface.

Tuning the model parameters so that the equilibrium position is reduced to below \SI{250}{\nano\meter} requires careful choice of the sign in $\vec F_\mathrm{el}$ (Eqn.~\ref{eq:el}). 
If using the negative sign, with very short distances, the force term gets negative (not visible in the logarithmic scale of Fig.~\ref{fig:force_distance}), hence permanently attracting the particle. The reduction of equilibrium height can happen, for instance, by using different particles with higher susceptibility, by increasing the strength of the external magnetic field in $z$-direction or by changing the electrostatic zeta potentials.

Reducing the gap to a few nm, $\vec F_\mathrm{vdW}$ is increased by more than 1000 times and has a significant influence on the height of the particle. 
If $|F_\mathrm{vdW}|+|F_\mathrm{gb}|+|F_{\mathrm{m,}z}| >|F_\mathrm{el}|$, the particle will move towards the polymer and adhere to its surface. 
Hence, a sufficiently strong repulsive electrostatic force and an equal sign for the zeta potentials of the particle and polymer layer are necessary to prevent this adherence. 
\subsubsection{Initial adherence of magnetic particles}\label{sssec:initialadherence}
In current experiments, a possible adherence of a particle to the polymer surface, permanent or temporary, cannot be seen. Using the optical defocusing-based 3D particle tracking method, only the relative change in particle height can be measured.
To incorporate a possible adherence into the simulation model, we artificially hold the particle in place of the equilibrium position and release it after a delay time. 

\begin{figure}[ht]
    \centering
\includegraphics[width=0.35\textwidth]{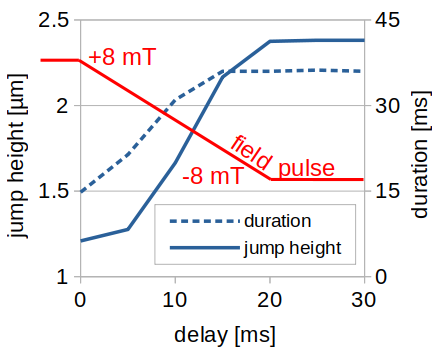}
    \caption{Jump height and duration of a magnetic particle with initial adherence. Delay is given from time point \SI{20}{\milli\second}, where the external field in $z$-direction starts to change from +\SI{8}{\milli\tesla} to \SI{-8}{\milli\tesla} (red line). At a delay of \SI{20}{\milli\second}, the field reaches its final value.}
    \label{fig:sticking}
\end{figure}

The $z$-component of the external magnetic field is changed from +\SI{8}{\milli\tesla} to \SI{-8}{\milli\tesla} in \SI{20}{\milli\second}; hence, depending on the delay, the particle will experience a different initial force. The duration and height of the jump depend on this force and are plotted against the artificial delay time, determined from the start of the external field switching, in Fig.~\ref{fig:sticking}.

Until about \SI{5}{\milli\second} of initial adherence, the height of the jump is only slightly increased. With longer adherence, both jump duration and height drastically increase until they reach a plateau at which the external field is completely switched. 
\subsubsection{Initial particle position and equilibrium time}\label{sssec:initialparticlepos}
\begin{figure}[ht]
    \centering
    \includegraphics[width=\linewidth]{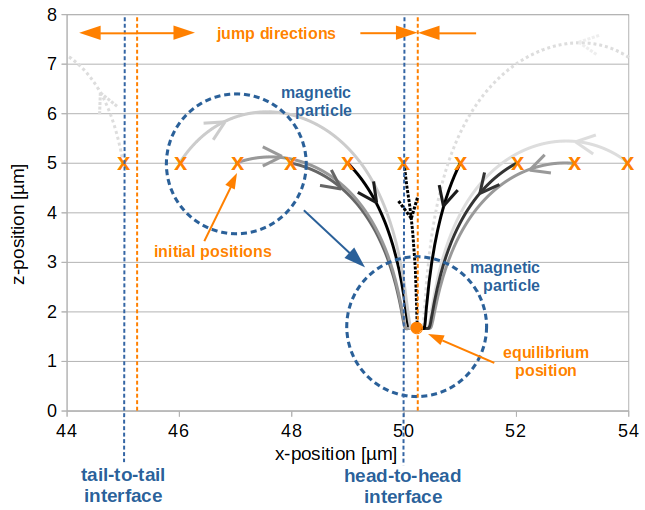}
    \caption{Several initial positions of magnetic particles and their trajectories to the equilibrium position are compared with constant $H_\mathrm{ext,z}$ = \SI{8}{\milli\tesla} and $H_\mathrm{ext,x}$ = \SI{1}{\milli\tesla}. At the orange dashed lines the jump direction is reversed. Locations of head-to-head and tail-to-tail interfaces are depicted with blue dashed lines.}
    \label{fig:initialPos}
\end{figure}

The initial position of the magnetic particle immersed in the fluid, when approaching the magnetic stray field landscape, is important for the resulting equilibrium position. In Fig.~\ref{fig:initialPos}, we compare particle trajectories for different initial positions at a height of \SI{5}{\micro\meter} above the stripe domain pattern. We investigate initial positions right above the tail-to-tail and head-to-head interfaces, as well as equally spaced intermediate positions with \SI{1}{\micro\meter} distance.


The direction of the particle trajectory is defined by the external field configuration and their relative position to the interfaces. With an initial positive external magnetic field $H_\mathrm{ext,z}$ the particle moves from tail-to-tail to head-to-head interface. The area between the orange dashed lines in Fig.~\ref{fig:initialPos} marks the jump directions. The additional constant magnetic field $H_\mathrm{ext,x}$ causes the equilibrium $x$-position to shift away from the head-to-head interface by about \SI{250}{\nano\meter}. The particle jump direction is reversed when the initial position approaches \SI{250}{\nano\meter} toward the tail-to-tail interface.
Compare also the discussion in Section~\ref{ssec:res_variation} and the large impact on the particle trajectory caused by the variation of $H_\mathrm{ext,x}$. Please note, that the pattern of particle trajectories is periodically repeated along the $x$-position, due to the underlying periodic domain pattern.
The duration for a particle to reach the equilibrium position can be very different. From an initial position \SI{5}{\micro\meter} above head-to-head it takes about \SI{23}{\milli\second} (black dashed line in Fig.~\ref{fig:initialPos}), whereas it takes more than \SI{350}{\milli\second} to reach the equilibrium position starting above tail-to-tail (light gray dashed line in Fig.~\ref{fig:initialPos}). The duration is further increased when the initial position is approaching the reverse point of the jump direction right next to tail-to-tail. 
Note, that an initial jumping event may cause an unexpected trajectory if the equilibrium time is not long enough and the equilibrium position cannot be reached in time. The same could happen with consecutive jumping events if triggered too fast. 
\newline


\section{\label{sec:conclusion}Conclusion}
A simulation of magnetic particle motion, immersed in a fluid and close to a polymer surface, activated by a dynamically transformed magnetic stray field landscape, follows multiple physical principles. These are fluid dynamics and the interaction with a rigid particle, gradient fields of the magnetic stray field and superposed external field contributions, surface charges repelling and attracting the particle according to electrostatic and van-der-Waals forces, and other minor contributions, like gravitation and Brownian motion. Each of the included physical principles is modeled by using a set of parameters. Understanding the effect of each parameter on the resulting trajectories of a particle allows us to control the movement and, further on, may offer the possibility to sense analytes (e.g., proteins) bound to the particles, which is crucial for medical applications. Here, we are focusing on parallel micron-sized stripe domains with a periodically alternating head-to-head/tail-to-tail magnetization configuration, which generates the necessary magnetic stray field landscape for the magnetic particle actuation.

In this work, we implemented all force contributions into the lattice Boltzmann framework of ESPResSo, which allows the dynamic interaction of the particle with the surrounding fluid. Repelling electrostatic and attracting magnetic forces lead the particle to a certain equilibrium position, whose absolute values cannot yet be measured in experiments. By applying a reversing external magnetic field, slightly bent with an additional field term, the particle can jump from a position above one shared interface of neighboring stripe domains (domain wall) to the next. 

The magnetic stray field in the simulations is obtained from analytical equations of cuboidal permanent magnets. 
For small distances, where the domain walls have a considerable impact, large-scale micromagnetic simulations are required, which currently are not yet possible due to size limitations in the computation. 

A single particle jump starts with a slow increase in height right after the external magnetic field begins to switch. This is caused by a reduced magnetic gradient force with respect to the repelling electrostatic force. After the field crosses zero, the particle magnetization is reversed, and the initial attraction to a head-to-head or tail-to-tail interface is now repelling the particle, which causes the actual jump. This jump could be visualized in 3D measurements, yet the slight increase in height right after the field started to switch was not observed due to the measurement uncertainty, leading to a certain threshold, after which a particle movement in vertical $z$-direction can be detected. This threshold is important to be included in the definition of a jump duration, otherwise, simulation and experiments cannot be compared.

Variations of the model parameters have shown that the magnetic properties can change the particle trajectories to a large extent, whereas electrostatic forces primarily influence the equilibrium position of a particle above the polymer spacer layer and its trajectory only slightly. This changes, when particles are moving closer to the polymer surface. Here, temporary sticking or even permanent adhesion can be caused by, for instance, van-der-Waals interactions. In current experiments, the particle equilibrium position is challenging to quantify; the simulation results provide, therefore, valuable predictions.

The most influential parameters found in this work are the magnetic susceptibility of the particles and the applied magnetic field 
parallel to the polymer surface and perpendicular to the stripe domains' long axis (compare Fig.~\ref{fig:model}).

The magnetic field applied parallel to the bilayer is necessary to control the direction of the particle jump. Slightly changing this field shifts the $x$-equilibrium position, thus changing the magnetic gradient force, which alters the particle trajectory. An increase in the field brings the equilibrium position closer to the interface of touching stripes with the effect of a higher jump that lasts longer. With reducing the field the effect is reversed.  

Due to larger deviations in literature on the magnetic properties of the used superparamagnetic particles, an accurate comparison of the simulation results with the experiment should be treated with caution. The nanoparticles immersed in the micron-sized particles may not be homogeneously distributed. Additionally, the applied magnetic fields from the Helmholtz coils need to be set carefully. Microscope heads and other equipment positioned close to the sample and, therefore, potentially influencing the magnetic fields should be used with special care.

Even though we investigated many parameters in our magnetic particle trajectory simulations, some further uncertainties need to be thoroughly examined. A deviation of experimental results from the simulation predictions is still present. An appropriate analysis of the height-dependent drag coefficients of spherical particles moving close to a flat surface might be able to explain this issue. In this work, we already considered a varying distance between the particle surface and the surface of the underlying substrate, however, the influence of this varied distance on the flow resistance originating from the surrounding fluid and acting on the particle is not reflected. A first attempt for implementing a height-dependent drag into the simulation is taken in the Supplementary Information, but must be further explored in future work. 

Changing the particle geometry will have major effects on the resulting trajectories as well. Increasing the time resolution of the 3D imaging experiment and also increasing the vertical spatial resolution may deliver important insights into the particle trajectories. Yet the theoretical insights gained in our study pave the way toward more advanced particle motion experiments, especially concerning analyte detection in medical Lab-on-a-chip systems. Our simulations reveal that the binding of an analyte to a particle, which comes with a change in its surface potential, may lead to a significant change in its equilibrium position above the chip surface. Detection with optical methods could present a simple yet efficient way to quantify this change in equilibrium position and, accordingly, the amount of bound analyte.


\section*{Acknowledgement}
This work was funded in whole or in part by the Austrian Science Fund (FWF) I 5712-N. We, additionally, acknowledge funding by the German Research Foundation (DFG) under the project numbers 514858524, 433501699, 361396165, and 361379292.



\section*{Data Availability}
All data generated or analyzed during this study are available from the corresponding author upon reasonable request.

\bibliographystyle{elsarticle-num-names}
\bibliography{references.bib}
\section{\label{Supp_Info}Supplementary Information}

\subsection{Height dependent drag coefficent}\label{ssec:heightdrag}
According to literature a particle experiences different drag forces far away and close to a rigid wall~\cite{happel2012low}. In case of a particle moving parallel or perpendicular to a solid wall, the fluidic drag force in Eqn.~\ref{eqn:stokes} can be multiplied with a correction factor $f_\mathrm{d,\parallel}$ and $f_\mathrm{d,\perp}$, respectively. In our model we have a polymer layer in $xy$-plane and the jump is initiated in $x$-direction, hence we obtain a corrected drag force 

\begin{align}
    \vec{F}_\mathrm{d,corr}=\vec{F}_\mathrm{d}\circ\vec{f}_\mathrm{d}=\vec{F}_\mathrm{d}\circ \begin{pmatrix} f_\mathrm{d,||}\\ 1\\ f_\mathrm{d,\perp}\end{pmatrix}
\label{eq:Fd_corr_mult}
\end{align}

with $\circ$ beeing the element-wise vector multiplication.

Please note, that both expressions for perpendicular and parallel movements were obtained from individual experiments. It is not clear if the expressions can be used together to account for the correct dynamics of a particle close-to-substrate motion. And it is not clear if the perpendicular correction is valid when the particle is moving away from the surface, because the experiment was performed only by approaching the surface. Here we test the correction terms individually and in combination and show their influence on the particle trajectory.



\begin{figure}
    \centering
    \includegraphics[width=0.3\textwidth]{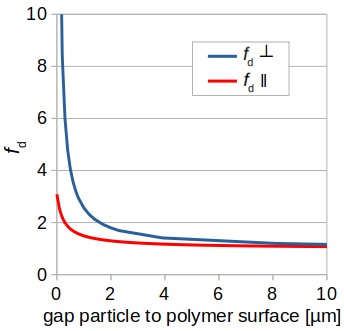}
    \caption{Fluidic drag correction coefficients $f_\mathrm{d,\parallel}$ and $f_\mathrm{d,\perp}$ are given for parallel and perpendicular motion (Equ.~\ref{eq:fd_parallel} and~\ref{eq:fd_perpendicular}), respectively, with given gap from particle surface to polymer surface.} 
    \label{fig:drag_correction}
\end{figure}

Faxen first described the motion of particles parallel to a single wall and found an expression for the parallel correction factor~\cite{faxen1923parallel}

\begin{align}
f_\mathrm{d,\parallel} = \left(1-\frac{9}{16}R+\frac{1}{8}R^3-\frac{45}{256}R^4-\frac{1}{16}R^5\right)^{-1}
\label{eq:fd_parallel}
\end{align}

with $R=r/l$, the particle radius $r$ and the distance $l$ between particle center to wall. With the correction factor, the drag force is increased up to a factor of 3 when approaching the wall, compared to a particle far away from the wall (Fig.~\ref{fig:drag_correction}).

In case of a particle falling towards a solid wall, Brenner described the perpendicular correction factor~\cite{brenner1961slow}

\begin{equation}
\begin{aligned}
f_\mathrm{d,\perp} = \frac{4}{3} \mathrm{sinh}~\alpha \sum_{n=1}^{\infty} \frac{n(n+1)}{(2n-1)(2n+3)} \\ 
\left(\frac{2 \mathrm{sinh}~(2n+1)\alpha + (2n+1)\mathrm{sinh}~2\alpha}{4 \mathrm{sinh}^2~(n+1/2)\alpha - (2n+1)^2\mathrm{sinh}^2~\alpha}-1\right)
\label{eq:fd_perpendicular}
\end{aligned}
\end{equation}

with $\alpha=\mathrm{cosh}^{-1}l/r$. At a gap from particle surface to polymer surface of about \SI{760}{\nano\meter} $f_\mathrm{d,\perp} \approx 3$, at about \SI{180}{\nano\meter} $f_\mathrm{d,\perp} \approx 9$ and at a gap approaching 0 $f_\mathrm{d,\perp}\rightarrow \infty$ (Fig.~\ref{fig:drag_correction}). 

In ESPResSo the drag force $\vec F_\mathrm{d}$ is implicitly handled via the 2-way coupling of the lattice Boltzmann fluid with the magnetic particle (compare Section~\ref{ssec:methods_LBM} and~\ref{ssec:res_calibration_drag}). In the current release of ESPResSo no height dependent drag is implemented, hence we have to add the missing force term $\vec{F^*_\mathrm{d}}$ to Equ.~\ref{eqn:stokes}.


\begin{align}
    \vec{F}_\mathrm{d,corr}=\vec{F}_\mathrm{d}\circ\vec{f}_\mathrm{d}=\vec{F}_\mathrm{d}+\vec{F^*_\mathrm{d}} 
\label{eq:Fd_corr_add}
\end{align}

Isolating $\vec{F^*_\mathrm{d}}$ yields 

\begin{align}
\vec{F^*_\mathrm{d}}=\vec{F_\mathrm{d}}\circ \left(\vec{f_\mathrm{d}}-\vec{1}\right)=\vec{F_\mathrm{d}}\circ\begin{pmatrix} f_\mathrm{d,||} -1\\ 0\\ f_\mathrm{d,\perp}-1\end{pmatrix}, 
\label{eq:Fd_corr_final}
\end{align}

which we implemented as an additional force term in ESPResSo. 





\begin{figure}
    \centering
    \includegraphics[width=0.4\textwidth]{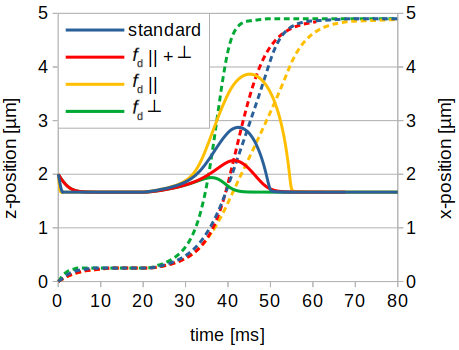}
    \caption{Comparison of magnetic particle trajectories with applied drag correction perpendicular, parallel to the polymer surface and a combination of both. $z$-positions are given with straight, $x$-positions with dashed lines. }
    \label{fig:jumps_drag_correction}
\end{figure}

In Fig.~\ref{fig:jumps_drag_correction} we compare the different contributions of the height dependent drag correction. 
The jump with the correction term $f_\mathrm{d,\perp}$ only is much lower and faster than the standard simulation.
Due to the increased drag in $z$-direction, the particle cannot reach a higher position. Applying only the parallel correction term $f_\mathrm{d,\parallel}$ the effect is reversed. The increased drag in $x$-direction decelerates the particle and causes an increased magnetic gradient force with the effect of a higher jump (compare initial adherence simulations in Section~\ref{ssec:res_uncertainties}). A combination of both drag corrections lowers the jump compared to the initial simulation. The jump duration, considering the \SI{500}{\nano\meter} $z$-threshold from the experiment, is reduced from \SI{14}{\milli\second} to \SI{5.2}{\milli\second}. 

\begin{figure*}[h]
\centering\includegraphics[width=\textwidth]{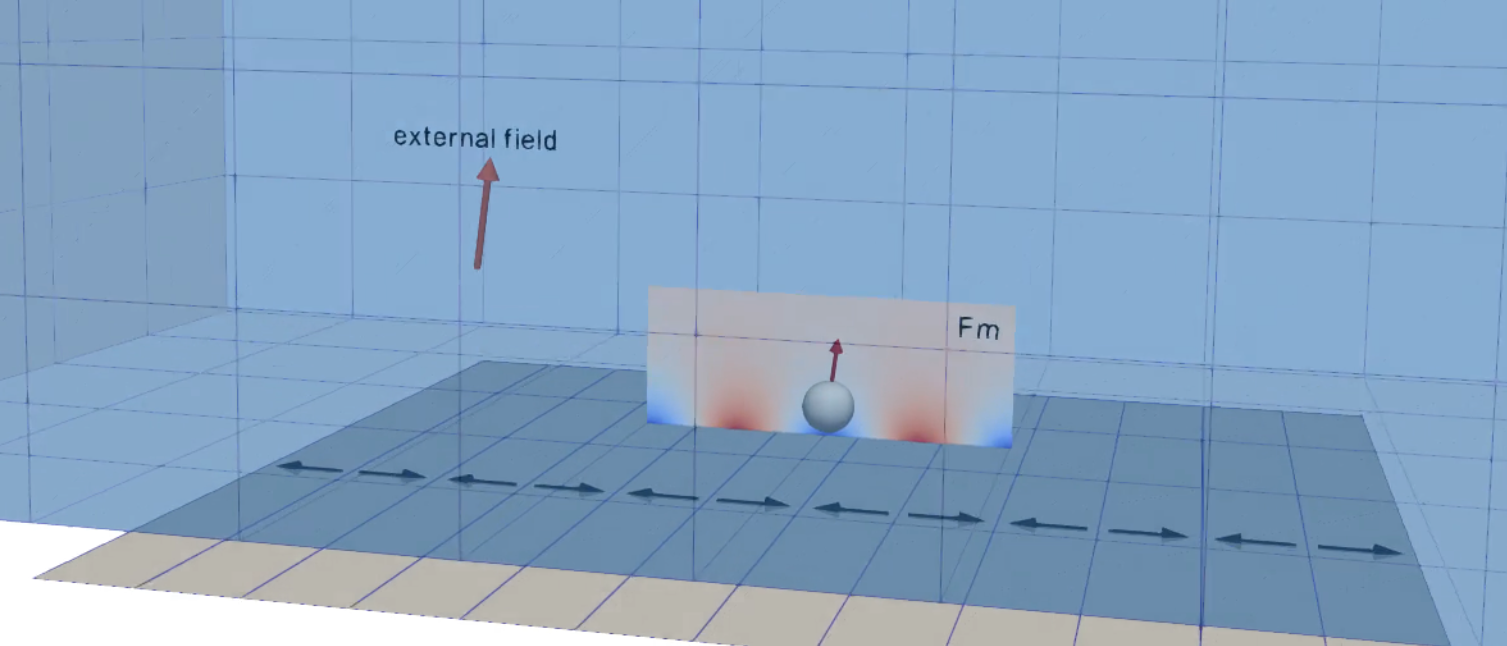}
    \caption{The large discretized box represents the lattice Boltzmann grid, whereas the underlying thin stripes represent the magnetic bilayer. In the video the black arrows indicate the magnetization directions of the magnetic bilayer. The non-magnetic polymer layer is not shown but fills the gap between bilayer and the fluid lattice. The $z$-component of the magnetic gradient force $F_\mathrm{m}$ acting on the particle is shown next to the particle with repelling (red) and attracting (blue) force values with respect to the magnetic bilayer. The scaled vector for external field and particle magnetization are shown on the left and directly at the particle, again colored with positive (red) and negative (blue) $z$-component. }
    \label{fig:video}
\end{figure*}
We assume, that the perpendicular force is less pronounced when the particle is leaving the surface, therefore the jumps including the term $f_\mathrm{d,\perp}$ may be possibly higher. Also, a combination of both correction terms is questionable, probably only a portion of each component is acting, depending on the particle moving direction.

    
    

\subsection{Video of a full simulation run}\label{ssec:video}

In Fig.~\ref{fig:video} we show a video frame of a full simulation run with the standard parameters of Tab.~\ref{tab:initial_parameters}. The video can be found in the supplementary materials.

\end{document}